\documentclass[twocolumn,showpacs,preprintnumbers,amsmath,amssymb]{revtex4}

\begin{document}

\title{Modified London Equation, Abrikosov-Like Vortices and Knot Solitons
        in Two-Gap Superconductors}
\author{Li-Da Zhang\footnote{Corresponding author.
                       Email: zhangld04@lzu.cn},
        Yi-Shi Duan, Yu-Xiao Liu}
\address{Institute of Theoretical Physics, Lanzhou University,\\
   Lanzhou 730000, P. R. China}

\begin{abstract}
We derive the exact modified London equation for the two-gap
superconductor, compare it with its single-gap counterpart. We
show that the vortices in the two-gap superconductor are soft (or
continuous) core vortices. In particular, we discuss the
topological structure of the finite energy vortices
(Abrikosov-like vortices), and find that they can be viewed as the
incarnation of the baby skyrmion stretched in the third direction.
Besides, we point out that the knot soliton in the two-gap
superconductor is the twisted Abrikosov-like vortex with its two
periodic ends connected smoothly. The relation between the
magnetic monopoles and the Abrikosov-like vortices is also
discussed briefly.

\end{abstract}



\maketitle

As an outstanding instance of the condensed matter systems with
several coexisting Bose condensates, the two-gap superconductor
(TGS) attracts wide interests, both theoretically and
experimentally \cite{NagamastuNature2001410}. The discovery of the
mapping between a two-flavor Ginzburg-Landau model and a version
of nonlinear $O(3)$ $\sigma$-model reveals the topological essence
of the TGS \cite{BabaevPRB200265}. Based on this mapping, many
topological solitons in the TGS, including vortices, knot
solitons, magnetic monopoles, etc, have been studied
\cite{BabaevPRB200265,BabaevPRL200289,BabaevNPB2004686,
JiangPRB200470}. In this paper, we derive the exact modified
London equation for the TGS, compare it with its single-gap
counterpart. We find the cores of vortices in the TGS are soft (or
continuous). This is distinct from the case in the single-gap
superconductor (SGS), where the Abrikosov vortices have hard (or
singular) cores. In particular, we discuss the topological
structure of the finite energy vortices (Abrikosov-like vortices),
and find that they can be viewed as the incarnation of the baby
skyrmion stretched in the third direction. Besides, we point out
that the knot soliton in the TGS is the twisted Abrikosov-like
vortex with its two periodic ends connected smoothly. The relation
between the magnetic monopoles and the Abrikosov-like vortices is
also discussed briefly.

We start by reviewing the mapping between a two-flavor GL model
and a version of nonlinear $O(3)$ $\sigma$-model. A TGS is
described by the two-flavor (denoted by $\alpha=1,2$)
Ginzburg-Landau free energy density
\cite{BabaevPRB200265,BabaevPRL200289}
\begin{eqnarray} \label{gl}
F&=&\frac{1}{2m_1} \left| \left( \nabla - i e {\bf A}\right)
    \Psi_1 \right|^2 + \frac{1}{2m_2}  \left| \left( \nabla
    - i e {\bf A}\right) \Psi_2 \right|^2 \nonumber \\
 & &+ \frac{{\bf B}^2}{8\pi} +{V} (|\Psi_{1,2}|^2)
    +\eta [\Psi_1^*\Psi_2+\Psi_2^*\Psi_1],
\end{eqnarray}
where $\Psi_\alpha=|\Psi_\alpha|e^{i\phi_\alpha}$,
${V}(|\Psi_{1,2}|^2)=-b_\alpha|\Psi_\alpha|^2+\frac{c_\alpha}{2}
|\Psi_\alpha|^4 $, and $\eta>0$ is a characteristic of the
interband Josephson coupling strength. Introduce new variables
$\rho ^2 = \frac12 \left( \frac{|\Psi_1|^2}{m_1}
+\frac{|\Psi_2|^2}{m_2} \right )$, ${\bf C}=\frac{i}{m_1 \rho^2}
[\Psi_1^*\nabla \Psi_1- \Psi_1 \nabla \Psi_1^*] +\frac{i }{m_2
\rho^2} [\Psi_2^*\nabla \Psi_2- \Psi_2 \nabla \Psi_2^*] + 4e{\bf
A}$, and ${\bf{n}}=(n_1,n_2,n_3)
=(\sin\theta\cos\gamma,\sin\theta\sin\gamma,\cos\theta)$, where
$\gamma=(\phi_2 -\phi_1)$ and
$[\cos\left(\frac{\theta}{2}\right),\sin\left(\frac{\theta}{2}\right)]=
[\frac{|\Psi_{1}|}{\sqrt{2m_1}\ \rho},
\frac{|\Psi_{2}|}{\sqrt{2m_2}\ \rho}]$. Then the original GL free
energy density (\ref{gl}) can be represented as
\cite{BabaevPRB200265,BabaevPRL200289}
\begin{eqnarray}\label{o3}
F&=& \frac{\rho^2}{4}(\nabla {\bf{n}} )^2 + (\nabla \rho)^2 +
\frac{\rho^2}{16} {\bf C}^2 + V(\rho , n_3 )+\rho^2 K n_1
\nonumber \\
& &+ \frac{1}{128\pi e^2} \left(\nabla\times{\bf
C}+\epsilon_{abc}n_a \nabla n_b \times \nabla n_c \right)^2
\end{eqnarray}
where $K\equiv 2\eta\sqrt{m_1m_2}$, $ V(\rho , n_3 )= A + B n_3 +
C n_3^2$, and the coefficients $A,B,C$ are given by $A=\rho^2 [
4c_1m_1^2 + 4c_2m_2^2 -b_1m_1 - b_2 m_2 ]$, $B= \rho^2 [ 8c_2m_2^2
- 8c_1m_1^2 -b_2m_2 + b_1 m_1]$, $C= 4\rho^2 [c_1m_1^2 +
c_2m_2^2]$. Then the potential term $V(\rho , n_3 )$ determines
the vacuum value of $n_3$ to be $\cos\theta_0\equiv
\left[\frac{N_1}{m_1}-\frac{N_2}{m_2}\right] \left[\frac{
N_1}{m_1} + \frac{N_2}{m_2}\right]^{-1}$, where
$N_\alpha=\langle|\Psi_{\alpha}|^2\rangle=b_\alpha/c_\alpha$.
Furthermore, taking account of the term $\rho^2 K n_1$, the vacuum
value of $\bf{n}$ is determined to be
${\bf{n}}_0=(-\sin\theta_0,0,\cos\theta_0)$. The models (\ref{gl})
and (\ref{o3}) have four characteristic length scales: condensate
coherence lengths $\xi_1$ and $\xi_2$, magnetic field penetration
length $\lambda=\frac{1}{e\rho}$ and the length scale associated
with the interband Josephson effect
\cite{BabaevPRL200289,BabaevNPB2004686}. For convenience, in the
following discussion, we assume that $\xi_1$ and $\xi_2$ have the
same order of magnitude and define the system coherence length
$\xi=max [\xi_1, \xi_2]$.

According to free energy density (\ref{o3}), the magnetic field in
the TGS is separated into two part: the contribution from ${\bf
C}$, which is equal to $\frac{1}{4e}\nabla\times{\bf C}$, and the
self-induced magnetic field $\tilde{\bf{B}}\equiv\frac{1}{4e}
\epsilon_{abc}n_a \nabla n_b \times \nabla n_c$, which is
originated from the nontrivial electromagnetic interaction between
the two condensates \cite{BabaevPRB200265}. At this point, we
might as well recall that the magnetic field in the SGS is also
separated into two part: the contribution from the supercurrent,
which is the counterpart of the above-mentioned contribution from
${\bf C}$, and the magnetic field with a $\delta$-function
distribution, which, as we will show below, is the counterpart of
the self-induced magnetic field $\tilde{\bf{B}}$. From the
expression of $\tilde{\bf{B}}$, we can see that this part of
magnetic field has a continuous distribution instead of the
singular distribution of its counterpart in the SGS. Besides, the
expression of $\tilde{\bf{B}}$ has an obvious topological meaning,
and therefore embodies the topological feature of the system.

For investigating the magnetic field distribution feature in the
TGS, we need the modified London for model (\ref{gl}) (or
equivalently model (\ref{o3})). For comparison with that in the
SGS, we will first review the derivation of the modified London
equation for the SGS which is usually written as
\cite{deGennes1966}
\begin{equation} \label{single0}
{\bf B}-\lambda^2\nabla^2{\bf B}
 =\Phi_0\sum_{k}\int d{\bf x}_{k} \delta^{3}({\bf x}-{\bf x}_{k}),
\end{equation}
where $\lambda$ is the London penetration depth,
$\Phi_0=\frac{2\pi}{e}$ is the standard flux quantum, and the line
integral is taken along the $k$-th vortex. During this review
process, some corrections to Eq. (\ref{single0}) will be added.
The supercurrent for the SGS is written as ${\bf J}=-\frac{ie}{2m}
[\Psi^*\nabla \Psi - \Psi \nabla \Psi^*] -\frac{e^2
}{m}|\Psi|^2{\bf A}$. Combining $\bf B=\nabla\times\bf A$,
$\nabla\times\nabla\times\bf B=\nabla\times(\nabla\cdot\bf
B)-\nabla^2\bf B$ as well as the Maxwell equations
$\nabla\times\bf B=4\pi\bf J$, $\nabla\cdot\bf B=0$, we find
\begin{eqnarray} \label{single1}
 {\bf B}&-&\frac{m}{4\pi e^2|\Psi|^2}\nabla^2{\bf B}+\frac{m}
{4\pi e^2} \nabla\frac{1}{|\Psi|^2}\times\nabla\times\bf B \nonumber\\
&=&-\frac{i}{2e}\nabla\times[\frac{\Psi^*}{|\Psi|^2}\nabla\Psi
-\frac{\Psi}{|\Psi|^2}\nabla\Psi^*].
\end{eqnarray}
Using the $\phi$-mapping method \cite{DuanNPB1998514}, we can
write the right-hand side (RHS) of Eq. (\ref{single1}) as
$-\frac{i\pi}{e}\delta(\Psi)\nabla\Psi^* \times\nabla\Psi$.
Expanding the $\delta$-function $\delta(\Psi)$, we arrive at our
modified London equation for the SGS:
\begin{eqnarray} \label{single2}
 {\bf B}&-&\frac{m}{4\pi e^2|\Psi|^2}\nabla^2{\bf B}+\frac{m}
{4\pi e^2} \nabla\frac{1}{|\Psi|^2}\times\nabla\times\bf B \nonumber\\
&=&\Phi_0\sum_{k}W_k\frac{d{\bf x}_{k}(s)}{ds}\int ds
\delta^{3}({\bf x}-{\bf x}_{k}(s)).
\end{eqnarray}
where $W_k$ is the flux quantum number of the $k$-th vortex, and
$s$ is the line parameter. Eq. (\ref{single2}) is different than
Eq. (\ref{single1}) mainly in the following aspects: (i) Eq.
(\ref{single2}) can describe the situation including the
multi-vortex, (ii) the London penetration depth in Eq.
(\ref{single2}) is a variant, and (iii) Eq. (\ref{single2})
includes an additional correction term, namely the third term in
the left-hand side of it. In spite of all these differences, Eq.
(\ref{single2}) retains the main feature of Eq. (\ref{single0}):
excluding the contribution from the supercurrent ${\bf J}$, the
magnetic field has a $\delta$-function distribution described by
the RHS of Eq. (\ref{single2}).

For the TGS, the supercurrent described in model (\ref{gl}) is
written as ${\bf J}=-\frac{ie}{2m_1} [\Psi_1^*\nabla \Psi_1 -
\Psi_1 \nabla \Psi_1^*] -\frac{ie}{2m_2} [\Psi_2^*\nabla \Psi_2 -
\Psi_2 \nabla \Psi_2^*] -2e^2 \rho^2{\bf A}$. As the above for the
SGS, we can get the following equation for the TGS:
\begin{eqnarray} \label{two1}
 {\bf B}&-&\frac{1}{8\pi e^2\rho^2}\nabla^2{\bf B}+\frac{1}
{8\pi e^2} \nabla\frac{1}{\rho^2}\times\nabla\times\bf B \nonumber\\
&=&-\frac{i}{4em_1}\nabla\times[\frac{\Psi_1^*}{\rho^2}\nabla\Psi_1
-\frac{\Psi_1}{\rho^2}\nabla\Psi_1^*]+(1\rightarrow2).
\end{eqnarray}
Then, using the $\phi$--mapping method as the above, we can write
the RHS of Eq. (\ref{two1}) as
\begin{eqnarray} \label{two1r}
-\frac{i}{4em_1}\nabla\frac{|\Psi_1|^2}{\rho^2}\times\Bigg[\frac{\Psi_1^*}
{|\Psi_1|^2}\nabla\Psi_1-\frac{\Psi_1}{|\Psi_1|^2}\nabla\Psi_1^*\Bigg]\nonumber \\
 -\frac{i\pi}{2em_1}\frac{|\Psi_1|^2}{\rho^2}\delta(\Psi_1)\nabla\Psi_1^*
\times\nabla\Psi_1+(1\rightarrow2).
\end{eqnarray}
Here we note that the term including $\delta(\Psi_1)$ and its
$\Psi_2$ counterpart vanish identically. This shows that the
singular part of the magnetic field, which is dominant in the SGS,
is replaced by the part of the magnetic field originated from the
continuous interaction between the two condensates. Furthermore,
from the definition of ${\bf{n}}$, Eq. (\ref{two1}) can be
rewritten in the following compact form:
\begin{eqnarray} \label{two2}
 {\bf B}&-&\frac{1}{8\pi e^2\rho^2}\nabla^2{\bf B}+\frac{1}
{8\pi e^2} \nabla\frac{1}{\rho^2}\times\nabla\times\bf B \nonumber\\
&=&\frac{1}{4e}\epsilon_{abc}n_a \nabla n_b \times \nabla n_c.
\end{eqnarray}
This expression is the exact modified London equation for the TGS.
Comparing Eq. (\ref{two2}) with Eq. (\ref{single2}), we find that
the self-induced magnetic field $\tilde{\bf{B}}$ is indeed the
counterpart of the singular magnetic field in the SGS.

Now we turn to the investigation of the topological structure of
the finite energy vortices in model (\ref{gl}) (or equivalently
model (\ref{o3})). In Ref. \cite{BabaevPRB200265}, Babaev
discussed various vortices in model (\ref{gl}). Among them, only
the vortex characterized by $\Delta( \phi_1 +\phi_2) \equiv\oint
d{\bf{l}} \cdot \nabla( \phi_1 +\phi_2)=4 \pi m$ and $\Delta\gamma
\equiv\oint d{\bf{l}}\cdot \nabla\gamma = 0$ (where we integrate
over a closed curve around the vortex core) has finite energy per
unit length \cite{BabaevPRL200289,BabaevNPB2004686}. Such a vortex
is an analog of the ordinary Abrikosov vortex in the SGS
characterized by $\frac{|\Psi|^2}{m}=\left(\frac{|\Psi_1|^2}{m_1}
+\frac{|\Psi_2|^2}{m_2}\right)$, because if both phases
$\phi_{1,2}$ change by $2\pi m$ around the its core, the vortex
will carry $m$ quanta of magnetic flux \cite{BabaevPRL200289}. In
the following, we will refer to such vortices as Abrikosov-like
vortices for short. In spite of the similarity between the
Abrikosov-like vortices and the ordinary Abrikosov vortices, they
have very different topological structures. To see this, let us
note the following two points. First, as we have said above, the
self-induced magnetic field distribution in the present system is
continuous instead of the singular distribution in the SGS. This
implies that the vortices in the present system are soft core
vortices while the Abrikosov vortices are the hard core vortices.
Second, if we note that an Abrikosov vortex has a vanishing
condensate at the center of its core, we may think that the
Abrikosov-like vortex has both vanishing condensates at the center
of its core. But this is not the case because at the points where
both $|\Psi_1|$ and $|\Psi_2|$ vanish, $\bf{n}$ can not be well
defined. Actually, around such a point, $\bf{n}$ has a
hedgehog-like distribution, which makes this point corresponding
to a magnetic monopole. Such monopoles in the TGS have been noted
by Jiang \cite{JiangPRB200470}. We will briefly discuss the
relation between these monopoles and the Abrikosov-like vortices
at the end of this paper.

To be specific, let us consider an Abrikosov-like vortex located
along $z$ axis. Because the vortex has finite energy per unit
length, ${\bf{n}}$ must tend to its vacuum value when the distance
away from the center of the vortex core extends to $\xi$. This
boundary condition compactifies $xy$ plane into $S^2$ and makes
${\bf{n}}$ a map: $S^2\mapsto S^2$. At this point, we can see that
the topological stability of the Abrikosov-like vortex originated
from $\pi_2(S^2)=Z$ rather than the Abelian topology
$\pi_1(S^1)=Z$, which guarantees the topological stability of the
ordinary Abrikosov vortex. In 2D, a topological soliton from
$\pi_2(S^2)=Z$ is known as a baby skyrmion. So we can say that an
Abrikosov-like vortex is an incarnation of a baby skyrmion
stretched in the third direction. In general case, we have
$N_1N_2\neq0$, or $\cos\theta_0\neq\pm1$. Then the curves formed
by the zeros of the two condensates are located in the soft core
of the vortex, but not at the center which corresponds to
$-{\bf{n}}_0$. If $N_2\rightarrow0$, or $\cos\theta_0
\rightarrow1$, the curves formed by the zeros of $\Psi_2$ and
$\Psi_1$ will tend to the boundary and center of the soft core
respectively, and the soft core itself will become hard gradually.

With the above topological analysis, we can construct a knot
soliton by twisting an Abrikosov-like vortex and connecting its
two periodic ends. The topological stability of the knot soliton
is guaranteed by the topology $\pi_3(S^2)=Z$. Due to the
self-induced feature of $\tilde{\bf{B}}$, the Abrikosov-like
vortex and the knot soliton can form in the TGS even in type-I
limit \cite{BabaevPRB200265}. According to Eq. (\ref{single2}), we
can make an analysis which concludes that the magnetic field
always decays in the magnetic field penetration length of the SGS.
As for the TGS, because $\tilde{\bf{B}}$ always disperses in the
scale of $\xi$, the similar analysis only applies to the magnetic
field originated from ${\bf C}$. This leads to the conclusion that
because of the interaction of the magnetic field itself, the size
of the knot soliton is of order $\xi(\lambda)$ in type-I(II)
limit.

Finally, we comment on the magnetic monopoles mentioned before.
From the topological structure of these monopoles, we find that
they must be connected by the Abrikosov-like vortices to form the
composite solitons. Due to the energy consideration, the monopoles
in such a composite soliton are supposed to present in
monopole-antimonopole pairs, and tend to annihilate. From this
analysis, we conjecture that a monopole-antimonopole pair
connected by an Abrikosov-like vortex may act as an ``instanton",
which could create and annihilate a knot soliton, and therefore
tunnel through the barrier between two topologically nonequivalent
field configurations. We will leave this subject to the future
studies.

\end{document}